%% file: main-R1.tex
\documentclass[11pt,twocolumns,article,nofontune]{IEEEtran}

\usepackage{url}
\usepackage[utf8]{inputenc}
\usepackage{xcolor}
\usepackage{amsmath}
\usepackage{amssymb}
\usepackage{xspace}
\usepackage{epsfig}
\usepackage{balance}

\usepackage[acronyms,nonumberlist,nopostdot,nomain,nogroupskip,acronymlists={hidden}]{glossaries}
\newglossary[algh]{hidden}{acrh}{acnh}{Hidden Acronyms}

\usepackage{booktabs}
\usepackage{tabularx}

\usepackage{tikz}
\usepackage{pgfplots}
\pgfplotsset{compat=newest}
\pgfplotsset{plot coordinates/math parser=false}
\newlength\fheight
\newlength\fwidth
\usetikzlibrary{plotmarks,patterns,decorations.pathreplacing,backgrounds,calc,arrows,arrows.meta,spy,matrix,scopes}
\usepgfplotslibrary{patchplots,groupplots}
\usepackage{tikzscale}
\usepackage[draft]{hyperref}

\newif\ifexttikz
\exttikzfalse

\ifexttikz
	\usetikzlibrary{external}
	\usepackage{fontspec}
\fi

\usepackage{multirow}

\usepackage[font=footnotesize]{subcaption}
\usepackage[font=footnotesize]{caption}

\usepackage{mathtools}
\usepackage[numbers,sort&compress]{natbib}

\usepackage{soul}


\input{acronyms.tex}
\glsdisablehyper
\input{myTikz.tex}

\definecolor{desireRed}{RGB}{230,57,60}%
\definecolor{darkPurple}{RGB}{59,31,43}%
\definecolor{springGreen}{RGB}{37,223,145}%
\definecolor{queenBlue}{RGB}{69,123,157}%
\definecolor{spaceCadet}{RGB}{29,53,87}%

\newcommand{\nearrt}{Near-RT \gls{ric}\xspace}
\newcommand{\nearrts}{Near-RT \glspl{ric}\xspace}
\newcommand{\nonrt}{Non-RT \gls{ric}\xspace}
\newcommand{\newrev}[1]{\textcolor{black}{{#1}}}

\usepackage{dblfloatfix}

\IEEEoverridecommandlockouts

\begin{document}

\title{Building An Open Architecture for AI-RAN Convergence in 6G}
\title{Beyond Connectivity: An Open Architecture for\\AI-RAN Convergence in 6G}


\author{\IEEEauthorblockN{Michele Polese, Niloofar Mohamadi, Salvatore D'Oro, Leonardo Bonati, Tommaso Melodia
\thanks{M. Polese, L. Bonati, and T. Melodia are with the Institute for the Wireless Internet of Things, Northeastern University, Boston, MA, USA. E-mail: \{m.polese, l.bonati, melodia\}@northeastern.edu. N. Mohamadi and S. D'Oro are with zTouch Networks, Inc. E-mail: \{niloofar.mohamadi, salvo\}@ztouchnet.com}}
\thanks{This article is based upon work partially supported by the NTIA PWSCIF under Award No. 25-60-IF054, the U.S. NSF under award CNS-2112471, and by OUSD(R\&E) through Army Research Laboratory Cooperative Agreement Number W911NF-24-2-0065. The views and conclusions contained in this document are those of the authors and should not be interpreted as representing the official policies, either expressed or implied, of the Army Research Laboratory or the U.S. Government. The U.S. Government is authorized to reproduce and distribute reprints for Government purposes notwithstanding any copyright notation herein.}}

\makeatletter
\patchcmd{\@maketitle}
  {\addvspace{0.5\baselineskip}\egroup}
  {\addvspace{-1.5\baselineskip}\egroup}
  {}
  {}
\makeatother

\flushbottom
\setlength{\parskip}{0ex plus0.1ex}

\maketitle
\glsunset{nr}
\glsunset{lte}
\glsunset{3gpp}
\glsunset{phy}
\glsunset{mac}

\begin{abstract}

Data-intensive \gls{ai} applications at the network edge demand a fundamental shift in \gls{ran} design, from merely consuming \gls{ai} for network optimization, to actively enabling distributed \gls{ai} workloads. 
This presents a significant opportunity for network operators to monetize \gls{ai} while leveraging existing infrastructure. To realize this vision, this article presents a novel converged O-RAN and AI-RAN architecture for unified orchestration and management of telecommunications and AI workloads on shared infrastructure.

The proposed architecture extends the Open RAN principles of modularity, disaggregation, and cloud-nativeness to support heterogeneous AI deployments. We introduce two key architectural innovations: (i) the AI-RAN Orchestrator, which extends the O-RAN \gls{smo} to enable integrated resource and allocation across \gls{ran} and \gls{ai} workloads; and (ii) AI-RAN sites that provide distributed edge \gls{ai} platforms with real-time processing capabilities. 
%
The proposed architecture enables flexible orchestration, meeting requirements for managing heterogeneous workloads at different time scales while maintaining open, standardized interfaces and multi-vendor interoperability.
\end{abstract}

\begin{picture}(0,0)(0,-420)
\put(0,0){
\put(0,0){\small This paper has been submitted to IEEE for publication. Copyright may be transferred without notice.}}
\end{picture}

\glsunset{nr}
\glsunset{lte}
\glsunset{3gpp}
\glsunset{phy}
\glsunset{mac}
\glsunset{oran}



\section{Introduction}



5G-Advanced and 6G networks are reshaping the role of the \gls{ran}, from a traditional communication substrate to a programmable, intelligent, ``sentient'', and compute-capable platform. With \newrev{emerging data-intensive and latency-sensitive use cases} such as \gls{xr}, \gls{isac}, and generative \gls{ai}, there is an opportunity to rethink \gls{ran} architecture not only as a consumer of \gls{ai} for internal optimization, but also as a distributed platform for hosting \gls{ai} workloads~\cite{kundu2025ai}. 
This transformation demands architectural innovation that supports the convergence of communication and AI computing within the network itself---simultaneously delivering connectivity and supporting distributed \gls{ai} inference and training demands. This dual capability is a foundational principle of the AI-RAN Alliance, a new telecom industry initiative dedicated to transforming networks through pervasive AI integration~\cite{AI-RAN_Whitepaper_2024}. 

In this paper, we design network resource orchestration and deployment workflows to \newrev{enable infrastructure} support \newrev{for} AI solutions (i) for the \gls{ran}, i.e., tasked with improving efficiency and operations; or (ii) on the \gls{ran}, i.e., leveraging the distributed compute infrastructure for inference, sensing, or training at the edge. The network infrastructure and orchestration abstraction layer supports flexibility, scalability, and interoperability to enable innovative AI-native next-generation wireless systems. At the same time, to enable innovation, this network abstraction and the underlying infrastructure need to embrace openness, standardized interfaces, disaggregated components and support for multi-vendor deployments.

\begin{figure*}[ht]
    \centering
    \includegraphics[width=.71\linewidth]{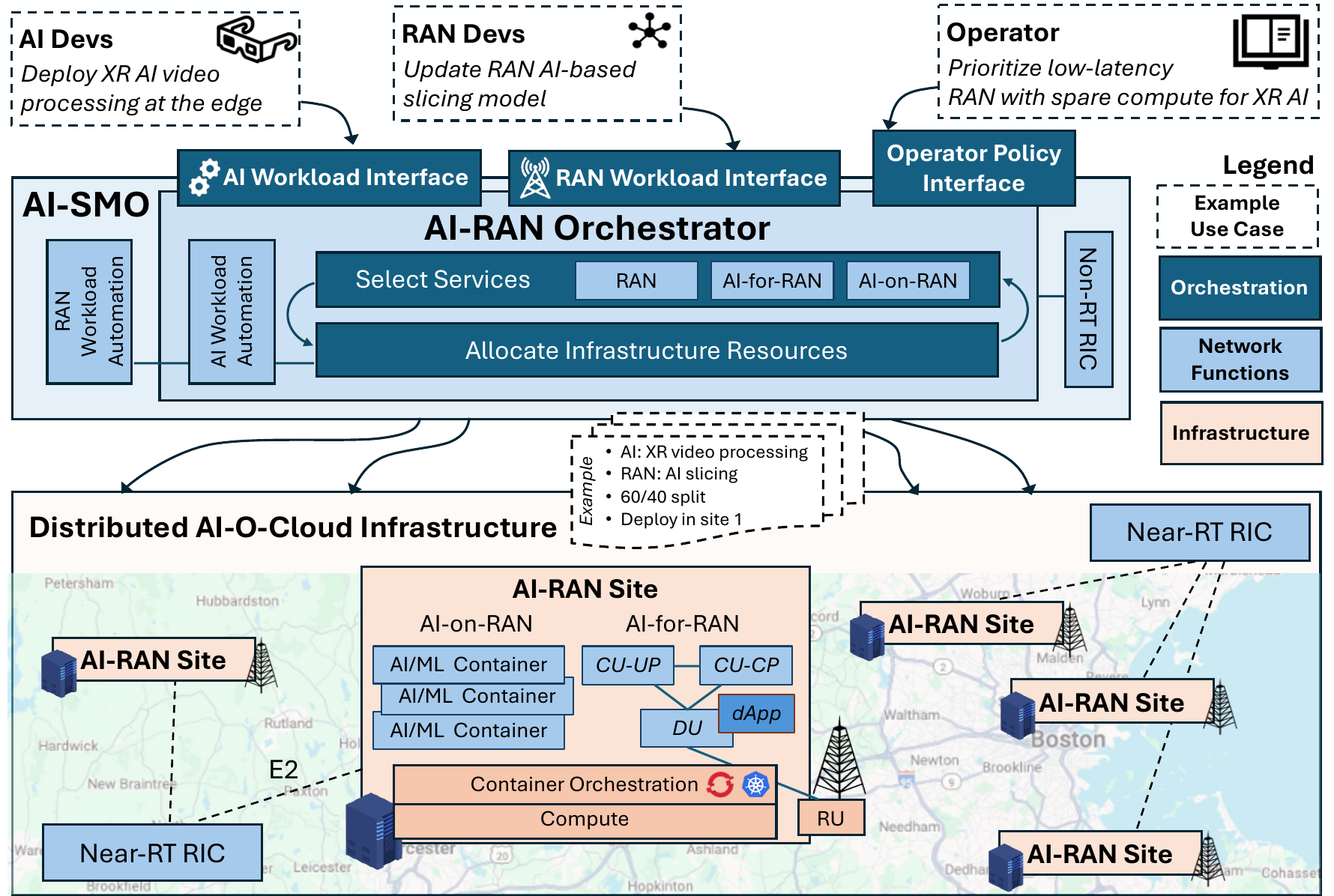}
    \caption{Architecture enabling AI and RAN coexistence, integrated within the proposed extension of the O-RAN framework. \newrev{The figure also includes an example of use case for AI-and-RAN coexistence. An application developer submits an AI-on-RAN workload to enable XR processing at the edge. In the meantime, a \gls{ran} vendor provides an updated AI-for-RAN slicing solution. The operator defines the policy that the infrastructure should follow to enable both services. The AI-RAN orchestrator translates the policy into resource allocation, and coordinates with the automation workflows to deploy the services in the AI-RAN sites.}}
    \label{fig:SMOS-arch-1}
\end{figure*}

\newrev{In this area, the prior work focused on edge \gls{ai}, open architectures, and early results on \gls{ai} and \gls{ran} orchestration.}
\newrev{Preliminary efforts in edge \gls{ai} deployment~\cite{duan2023distributed} did not lead to a cohesive architectural framework} that can effectively manage the complex interplay between \gls{ai} workloads and \gls{ran} operations at scale. The \gls{oran} ALLIANCE \newrev{introduced intelligent control loops to optimize \gls{ran} performance~\cite{polese2023understanding}. 
However, its architecture, with} open, interoperable interfaces, a \gls{smo} and the \glspl{ric}, \newrev{does not support generic AI workloads on} a shared \gls{ran} infrastructure.
As discussed above, this is \newrev{instead a focus area for} the AI-RAN Alliance. Recent literature has also addressed \gls{ai} and \gls{ran} orchestration~\cite{ananthanarayanan2024distributed,li2025rethinking,kundu2025ai,shah2025interplay,hung2025dmmp}. Specifically,~\cite{ananthanarayanan2024distributed} \newrev{and \cite{li2025rethinking}} discuss \gls{ai} operations in the cloud, edge, and also \gls{ran}, \newrev{but do not propose an architecture for orchestration}.
\cite{shah2025interplay} proposes a converged AI-and-O-RAN architecture \newrev{but the end-to-end orchestrator only interacts with} the \nearrt programmable logic units (xApps), thus limiting the scope of services that can be managed and deployed.
\newrev{\cite{hung2025dmmp} discusses AI-RAN \gls{ai}} operational workflows for AI-RAN, \newrev{but does not consider architectural elements}. \cite{kundu2025ai} \newrev{presents an infrastructure design that jointly supports} \gls{ran} and \gls{ai} workloads, \newrev{without considering AI-RAN orchestration}. 
\newrev{Therefore}, a comprehensive, end-to-end, and generalized architecture enabling AI and RAN coexistence following O-RAN management and orchestration principles is still missing.

\noindent\textbf{Contributions.} This paper takes a decisive step in this direction by proposing a converged O-RAN and AI-RAN architecture, where the O-RAN framework is extended to support the orchestration and management of AI workloads, as shown in Fig.~\ref{fig:SMOS-arch-1}.
By building upon the foundational principles of modularity, disaggregation, and cloud-nativeness of the existing \gls{oran} architecture, this work demonstrates how \gls{ai} integration can be achieved with minimal disruption and without increasing system complexity. The proposed design extends the \gls{ran} compute infrastructure to the \gls{ai} datacenter domain, introducing new monetization opportunities for public and private operators.
Specifically, we make the following key contributions:

\noindent $\bullet$ {\bf Gap Analysis.}  We review  AI-RAN activities and discuss the technical and operational challenges to achieve seamless AI-RAN and O-RAN integration.

\noindent $\bullet$ {\bf Unified Architecture Design.} We present a new {comprehensive architecture} that enables AI and RAN workloads to share infrastructure while maintaining performance isolation and service quality, shown in Fig.~\ref{fig:SMOS-arch-1}. 
We also identify and address the {orchestration requirements for managing heterogeneous AI and RAN workloads} at different time scales. 
%
Specifically, we identify and propose two key architectural components. (i) The \textit{AI-RAN Orchestrator} (Fig.~\ref{fig:SMOS-arch-1}, top) extends the O-RAN \gls{smo} framework for unified resource allocation, policy management, and workload coordination across AI and RAN services. (ii) The \textit{AI-RAN Sites} augment the O-RAN O-Cloud concept to further support AI within a common RAN infrastructure, enabling a distributed edge AI platform with real-time processing capabilities.

\noindent $\bullet$ {\bf Flexible Orchestration.} We define comprehensive procedures for deploying AI workloads on shared RAN infrastructure, introducing two distinct workflows for non-real-time batch and latency-critical deployments.


This work establishes {\em an architectural blueprint for next generation intelligent radio access networks} that go beyond traditional connectivity boundaries, enabling simultaneous delivery of telecommunications services and distributed edge AI capabilities on shared infrastructure.

\section{Toward AI-RAN}
\label{sec:ai-ran}

This section reviews the key elements of each AI-RAN Alliance working group (AI-for-RAN, AI-on-RAN, AI-and-RAN) and the O-RAN ALLIANCE architecture.
\subsection{AI-RAN Principles}
\label{sec:principles}

\textbf{AI-for-RAN.} 
The first pillar, AI-for-RAN, applies \gls{ai} to enhance network automation, efficiency, and adaptability \cite{li2025rethinking}. 
AI-for-RAN enhances RAN management through data-driven decision-making \newrev{and \gls{dsp}}, improving performance, reliability, and cost efficiency~\cite{AI-RAN_Whitepaper_2024}. This enables intelligence and self-optimization in existing RAN functions~\cite{AI-RAN_Whitepaper_2024,polese2023understanding}.

\textbf{AI-on-RAN.}
Conversely, AI-on-RAN enables \gls{ai} and generative \gls{ai} applications to execute on the \gls{ran} infrastructure. These applications may or may not be related to cellular systems, but leverage spare compute capacity and the geographically distributed nature of the \gls{ran}. \gls{ran} deployments are over-provisioned for peak loads, therefore, by leveraging \gls{ai} during low \gls{ran} activity periods, operators can maximize infrastructure efficiency without compromising connectivity  \cite{ananthanarayanan2024distributed}. 
By embedding \gls{ai} workloads within the \gls{ran}, AI-on-RAN brings AI close to end-users and thus reduces the latency and bandwidth overhead for transmission to centralized data centers. \newrev{This makes AI-on-RAN} ideal for latency-sensitive applications that require real-time inference and decision making. \newrev{It also helps distributing workloads from large data centers to the edge.}
Examples of AI-on-RAN workloads can include distributed AI training (e.g., federated learning), video processing and computer vision applications, or edge generative \gls{ai} solutions (e.g., a low-latency edge instance of a \gls{llm}, \newrev{chatbots, AI agents}), among others.

\textbf{AI-and-RAN.}
To enable infrastructure sharing across AI applications and \gls{ran} workloads, AI-and-RAN focuses on the coexistence of heterogeneous workloads. 
Indeed, \gls{ai} and \gls{ran} workloads differ in terms of reliability, availability, and performance requirements. The \gls{ran} \gls{dsp} represents a non-elastic workload, where tasks need to be executed within specific timing constraints to ensure intended functionality of the protocol stack (e.g., for channel coding/decoding, scheduling). At the same time, processing for a specific configuration (bandwidth, number of MIMO streams) requires a fixed and predictable amount of resources. \gls{ai} and generative \gls{ai} inference or training solutions are more elastic, provided that enough resources are allocated to load the data structures associated with the AI models. 


\textbf{AI-RAN Orchestration Requirements.} 
Considering these requirements, the orchestration of AI-for/on-RAN requires (i) a holistic accounting of the underlying compute and networking resources, so as to dynamically track infrastructure availability; (ii) northbound \glspl{api} to AI and RAN developers, to expose available capacity and to submit workloads to the system, thus enabling infrastructure monetization; (iii) an interface for operator requirements/intents (e.g., priority level for \gls{ran} workloads, maximum latency for AI inference); (iv) a scheduling strategy that combines requests, intents, and available resources into actionable policies or allocations; and (v) a southbound interface for AI and RAN configuration and deployment. This also allows for preemption for time-sensitive \gls{ran} workloads. Figure~\ref{fig:SMOS-arch-1} captures these requirements in the end-to-end architecture.

\subsection{\newrev{A Primer on Open RAN}}

The AI-RAN Alliance does not develop specifications or standards, \newrev{but builds on and influences those within 3GPP and the O-RAN ALLIANCE.}
%
The latter has driven the design and development of networked systems that abide by the principles of openness, programmability, interoperability, and intelligent closed-loop control. \newrev{A complete overview can be found in~\cite{polese2023understanding}.} 

\newrev{The O-RAN architecture is based on a softwarized}, disaggregated \gls{ran}, with atomic network functions that are deployed on the O-Cloud, a virtualization platform for cellular systems. \newrev{Through open interfaces, the} \glspl{ric} connect the \gls{ran} \newrev{to} custom intelligence in xApps (for the \nearrt) and rApps (for the \nonrt). \newrev{The} \gls{smo} provides management and orchestration across the whole network. \newrev{The O-RAN architecture already features pipelines and workflows for AI-for-RAN.}

The \gls{smo} and \nonrt concern network management in the time scale of seconds or more, for \gls{ran} services and \gls{ran} performance/efficiency, respectively. The latest O-RAN architecture decouples them in a service-based framework, where the \gls{smo} combines multiple independent services communicating through the standardized \gls{smos} interface. The O2 interface connects the \gls{smo} with the O-Cloud for infrastructure management and workload deployment~\cite{oran-wg6-o-cloud}.

\vspace{-.2cm}
\section{Challenges and Open Issues Toward AI-RAN and O-RAN Convergence}
\label{sec:challenges}

\newrev{The} cloud-native architecture of Open \gls{ran} can streamline \newrev{AI-RAN} innovation. \newrev{Open} interfaces and service composition \newrev{allow onboarding} software, workloads, and workflows from multiple vendors, providers, and users. The open ecosystem fosters multi-vendor collaboration, ensuring that \gls{ai} algorithms can be developed and deployed across diverse hardware and software platforms without vendor lock-in and with well-defined procedures. 
In addition, building on an existing architecture makes it possible to avoid unnecessary complexity, extending \gls{oran} components to also manage \gls{ai} workloads rather than introducing additional orchestration solutions. 

\newrev{Effective AI-and-RAN orchestration and integration with O-RAN system, however, requires addressing several challenges, which we discuss next.}

\textbf{Architecture and \newrev{Orchestration} Workflows Challenges.} AI-RAN coexistence requires an evolution of the 3GPP and O-RAN architectures. 
The end-to-end management system requires an entity responsible for AI-and-RAN coexistence, i.e., the orchestrator introduced in Fig.~\ref{fig:SMOS-arch-1} and in Sec.~\ref{sec:ai-ran}. \newrev{Challenges include} understanding what are \newrev{effective control} and data exposure abstractions, which is network host for such functionalities, and if orchestration requires a hierarchical approach. 

From a workflow point of view, there is a need for telemetry, control, and deployment pipelines, extending the O-RAN interfaces to define the southbound and northbound interfaces of the orchestrator (Fig.~\ref{fig:SMOS-arch-1}).
%
Southbound interfaces should support continuous telemetry collection, preprocessing, and delivery from distributed AI-RAN components, accommodating latency requirements for monitoring, inference, or control applications. This unlocks the unique opportunity to systematically create large-scale datasets.
Finally, what \glspl{kpm} are better suited to support orchestrator tasks is an open question.

Finally, the definition of AI-RAN-specific procedures for \gls{ai} lifecycle management implies regulating access to RAN resources, orchestrating deployment across nodes, and ensuring compliance with the operator's policies. To do so, current \gls{oran} interfaces (e.g., O1, O2, E2, A1) require targeted modifications to support the dual operation of RAN control loops and \gls{ai} workflows. 

\textbf{Orchestration Challenges.} An AI-RAN system (\newrev{as shown in Fig.~\ref{fig:SMOS-arch-1}}) combines compute with different \gls{ai} accelerators (e.g., \glspl{gpu}), storage, networking elements, radios, and spectrum. \newrev{The orchestrator needs to account for such heterogeneity} when allocating resources. \newrev{Additional compute can be available in the core network, which however does not have the same distributed nature as the \gls{ran}.} 
RAN nodes exhibit diverse processing capabilities, also based on deployment location, \newrev{which} can lead to inefficiencies in workload distribution \cite{kundu2025ai, liu2022cloud}. 
The heterogeneity also extends to the workloads. \newrev{As} discussed in Sec.~\ref{sec:ai-ran}, \gls{ai} and \gls{ran} often have different workload patterns and requirements. 
\gls{ai} itself can range from lightweight inference for real-time network optimization or value-added services to high-intensity model training. Moreover, the unpredictability of \gls{ai} workloads, and in particular that of large-scale training tasks, can introduce latency spikes that may negatively impact \gls{ran} performance \cite{kundu2023hardware, 3gpp_TR_38.843}. This complicates both monitoring, which needs to account for real-time variations in the workload intensity, and orchestration, which needs to combine multi-scale and proactive coexistence approaches. \newrev{Finally, orchestration workflows can extend beyond the \gls{ran}, toward core network functions that contribute to the \gls{ai} operations.}

\textbf{Resource Allocation Challenges.} The resource allocation problem must consider operators' priorities, the multi-dimensionality of the problem (encompassing a distributed infrastructure across different geographical locations with heterogeneous capabilities), and potentially a hierarchical control structure with varying granularity, from a centralized orchestrator to distributed components. The orchestrator input also comes from different stakeholders (e.g., AI developers, RAN vendors, operators), whose orthogonal requirements may need to be harmonized and reconciled. \newrev{Moreover, operators may impose different \gls{sla} to AI and RAN. These may change across sites and time and need to be accounted for when allocating resources.}
\newrev{Therefore, while solutions can be inspired by prior work on scheduling and slicing, they also need to take into account the complexity of the input, output, and dimensionality of the infrastructure partitioning problem.}

\textbf{Use Case, \newrev{Monetization, and Energy Efficiency} Challenges.} The AI-RAN orchestrator enables monetization strategies for next-generation wireless by identifying trade-offs, prioritizing, and provisioning strategies to satisfy both AI and RAN tenants. \newrev{To successfully cater to AI-on-RAN stakeholders, the orchestrator needs to align with AI and cloud best practices}. 

The coexistence and monetization challenge also extends to energy efficiency. With \gls{ran} energy consumption being the highest operational expenditure for operators to date, it is important to adopt an energy-efficient and opportunistic approach to the acceleration of \gls{ai}, \newrev{and to account for the impact of energy in the total costs of the AI-RAN infrastructure.} AI models are usually associated with high computational demands, in contrast to recent trends for low-power \gls{ran} components. Hardware acceleration offers potential mitigation but requires careful scheduling to ensure energy efficiency~\cite{kundu2023hardware} \newrev{while accounting for non-linearity and multi-dimensionality of the AI accelerators power consumption as a function of processor load, memory utilization, and data transfer}.

Integrating third-party AI workloads also presents challenges to security, privacy, and access management. The orchestrator \glspl{api} and the end-to-end control and infrastructure management pipelines need to ensure accountability and tracking of security roles and privileges. 
Deploying \gls{ai} models or workloads with vulnerabilities in the software stack could affect the stability of the \gls{ran}. Such attacks could lead to network congestion, degraded service quality, or large-scale outages \cite{khan2023ai}.

In the next section, we introduce an architecture designed to address these challenges while adhering to the design principle of minimizing system complexity, and considering factors such as privacy and security.

\section{Converged AI-RAN and O-RAN Architecture}
\label{sec:architecture}

\begin{figure}[t]
    \centering
    \includegraphics[width=.95\linewidth]{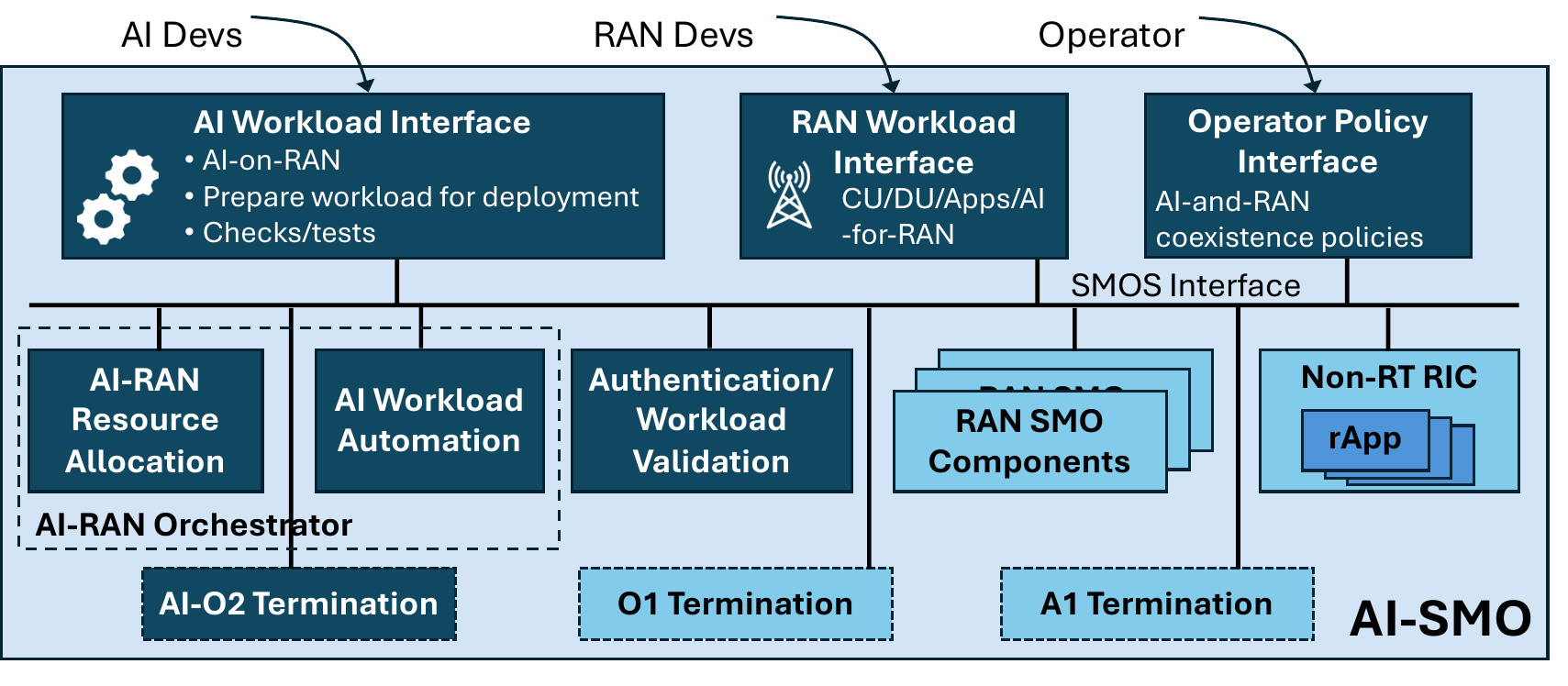}
    \caption{Extension of the O-RAN \gls{smo} to support AI-RAN orchestration. The components in \newrev{light blue}, together with the \gls{smos}, are part of the O-RAN \gls{smo} architecture. Components in \newrev{dark blue} represent extensions to accommodate AI-RAN requirements.}
    \label{fig:ai-smo}
\end{figure}

Building on the architecture in Fig.~\ref{fig:SMOS-arch-1} and the requirements identified above, Figs.~\ref{fig:ai-smo} and~\ref{fig:ai-ran-site} detail the specific AI-RAN extensions to O-RAN components and their interfaces. This converged architecture has been designed to enable two concurrent workflows for the orchestration, deployment, and execution of \gls{ai} workloads, i.e., batch and real-time. 
In batch mode, \gls{ai} tasks are submitted to the centralized orchestrator, and distributed when/where compute capacity is available. This can be used, for example, for AI training, or general AI processing that requires acceleration on \gls{gpu} but does not have constraints on the deployment location and timing of the deployment. In real-time mode, \gls{ai} developers submit jobs for direct execution to a specific cluster (or geographical area of interest) and the request will be accommodated by the orchestrator according to resource availability. 

\vspace{-.3cm}
\subsection{\newrev{AI-SMO and AI-RAN Site}} 

\newrev{\textbf{AI-SMO.}} Figure~\ref{fig:ai-smo} illustrates the architectural extension of the \gls{smo} toward an AI-enabled SMO (AI-SMO). Besides O-RAN \gls{smo} components connected through the \gls{smos} bus, a set of new micro-services covers the following functionalities: resource allocation, \gls{ai} workload automation, authentication and workload validation for the \gls{ai} components, and exposure of interfaces to interact with the AI-RAN elements. 
By embedding the AI-RAN orchestrator in the \gls{smo}, we provide the orchestrator with (i) access to granular monitoring data on the infrastructure already collected through the O-RAN O1 and O2 interfaces; and (ii) a global, centralized viewpoint on the distributed \gls{ran} infrastructure that enables a holistic approach to allocation of resources to either \gls{ai} or \gls{ran} workloads. 
The workload validation and \gls{ai} user authentication \newrev{provide access control with an} authorization process that vets users and \gls{ai} tasks before they are admitted to the system. 
It validates the identity of \gls{ai} users through the northbound interface to handle requests for training or inference execution. 
Upon successful validation, the \newrev{service} issues an authorization token granting access to the appropriate resources. \newrev{The authentication mechanism can be tailored to the specific deployment operational constraints, thanks to the disaggregated design of the AI-SMO.}
Then, upon availability of resources determined by the orchestrator, \gls{ai} task requests are forwarded to the \gls{ai} workloads automation component, which triggers their deployment to the AI-RAN infrastructure sites. 
This component operates in parallel to equivalent services for \gls{ran} workloads, including the \gls{nfo} and \gls{focom} services, with which it coordinates through the \gls{smos}.
Batch AI and RAN workloads are deployed to the selected AI-RAN edge sites using an extension of the O2 interface (AI-O2), which also communicates to the edge sites that resource allocation schedules or policies defined by the AI-RAN Orchestration service. \gls{qos} policies for AI-O2 traffic can further help by prioritizing \gls{ran} workloads over less time-sensitive \gls{ai} tasks. \newrev{Finally, operators can dynamically express geographically and temporally customized policies that determine AI and RAN sharing strategies.}



\begin{figure}[t]
    \centering
    \includegraphics[width=.95\linewidth]{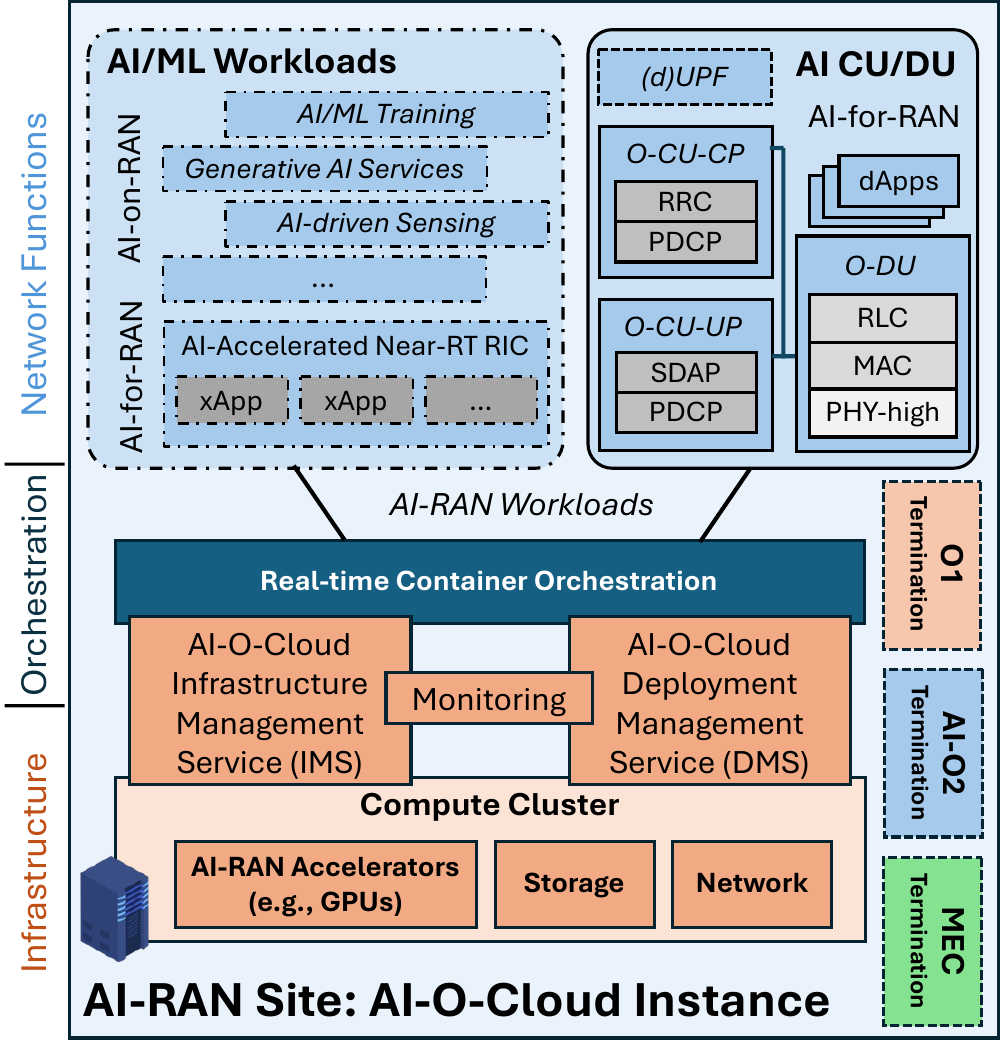}
    \caption{Architecture for an edge cloud site that supports AI-RAN solutions on a compute cluster managed by the AI-O-Cloud. \newrev{The figure shows the AI workloads and RAN workloads (top), the orchestration components (middle), and the infrastructure and interfaces (bottom)}.}
    \label{fig:ai-ran-site}
\end{figure}


\newrev{\textbf{AI-RAN Site.}} At the other end of the AI-O2 interface, the AI-RAN edge site, shown in Fig.~\ref{fig:ai-ran-site}, is responsible for the granular, container-level service orchestration. Here, the O-Cloud extends into an AI-ready version that can manage containerized workloads for AI and RAN solutions, \newrev{and also potentially distributed core components such as a local \gls{upf}}. Figure~\ref{fig:ai-ran-site} also shows the taxonomy of workloads that can be executed on AI-RAN sites. \newrev{Besides AI-driven \glspl{cu} and \glspl{du}, AI-for-RAN solutions include, for example, dApps~\cite{lacava2025dapps}, real-time programmable components that extend the O-RAN architecture to provide intelligence within the protocol stack}. An additional AI-for-RAN category, outside of the RAN CUs/DUs, is represented by AI-accelerated \nearrts, which can be hosted at the edge to satisfy near-real-time inference and control constraints. AI-on-RAN solutions can include training for AI models, generative AI services (e.g., chatbots), sensing based on AI, and other edge inference tasks, among others.

The AI-RAN site is equipped with compute (including AI accelerators, e.g., \glspl{gpu}), storage, and networking resources. \newrev{These are managed by} a container orchestration platform (e.g., Kubernetes or Red Hat OpenShift), \newrev{which also provides tenant's workload isolation via namespaces}. Both of these components are managed by the AI-O-Cloud \gls{ims} and \gls{dms}. These extend O-RAN services that terminate the O2 interface for infrastructure configuration and workload lifecycle management, respectively. In the AI-O-Cloud, they also translate policies and schedules determined by the \gls{smo} into AI infrastructure configurations and AI workload deployment, while enabling coexistence with the \gls{ran}.

%
\subsection{\newrev{Scalability and Real-time Deployment Workflow}}

\begin{figure}[t]
    \centering
    \includegraphics[width=.95\linewidth]{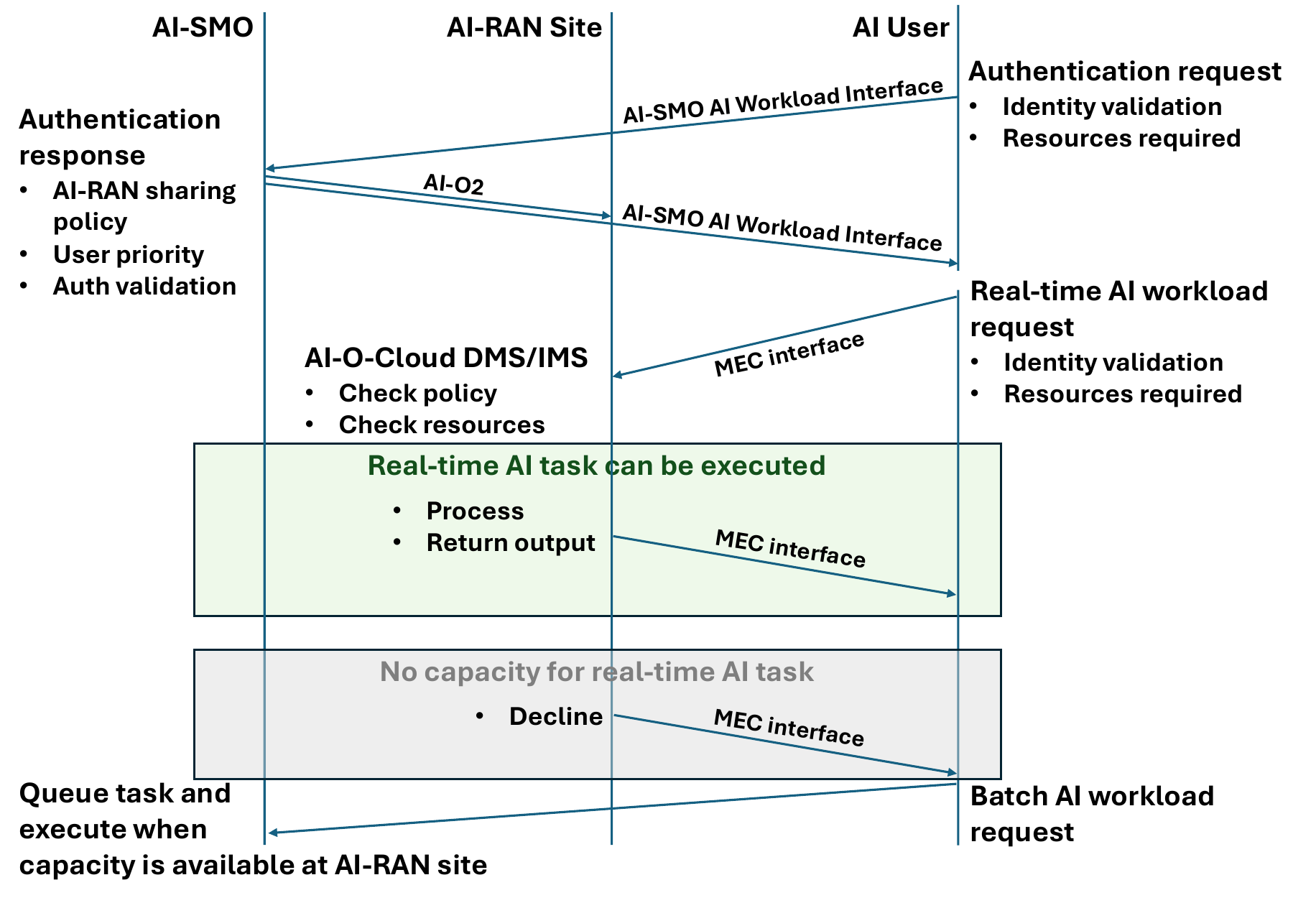}
    \caption{Procedure for the authentication and deployment of real-time AI tasks on the AI-RAN infrastructure.}
    \label{fig:workflow}
\end{figure}

\newrev{The service-based nature of the AI-SMO favors scalability, with different components that can be scaled to handle increasing workloads submissions or large number of sites. At the same time, the AI batch submission workflow and specific instances of optimization for the resource allocation may introduce delays when applied to large-scale systems. Further, specific AI-on-RAN task may require low-latency deployments. Examples include AI inference services that are dynamically offloaded by or need to follow highly mobile users in a network, e.g., connected autonomous vehicles. Public safety sensing applications may need rapid deployment in response to an emergency. For this reason,} in Fig.~\ref{fig:workflow}, we propose a procedure for the real-time deployment of AI workloads. It combines (i) pre-authentication with the AI-SMO, with an indication of computing requirements at specific AI-RAN sites; (ii) the definition of a resource sharing policy in the AI-RAN Orchestrator, which is then communicated to the edge site through AI-O2; (iii) the use of a \gls{mec} interface, adopting the ETSI standards for edge computing, by the pre-authenticated users to submit real-time AI workloads to a specific AI-RAN site; and (iv) involvement of the \gls{dms}, which allocates resources and deploys the AI services with minimal latency, as long as the amount of resources required is compatible with the policy defined by the AI-RAN Orchestrator. If not, the AI task is not admitted, and the process resumes at the AI-SMO, where the user can explore options to increase priority of the task (e.g., through different monetization tiers) or submit it as a batch workload or to a different AI-RAN site. 

\subsection{Profiling the AI-RAN Site}



\newrev{We profile the feasibility of the approach with a prototype of the AI-RAN site, focusing on AI and RAN coexistence and on their deployment latency. The results are based on multiple, independent experiments executed on the Northeastern University X5G testbed~\cite{villa2025x5g}, which implements a private 5G network with multiple commercial O-RAN \glspl{ru} and fronthaul infrastructure. We integrate in X5G a single-node OpenShift cluster with a Gigabyte server with an x86 CPU and an NVIDIA A100 GPU. For the \gls{ran} software, we use the NVIDIA Aerial GPU-accelerated physical layer and OpenAirInterface. For \gls{ai} workloads, we use different Ollama-based \glspl{llm} (processing prompts), and a \gls{cnn} or a ResNet model trained live on CIFAR-100. The AI-SMO configures the AI-RAN site \gls{gpu} with \gls{mig}, an NVIDIA technology for resource partitioning. For these experiments, the GPU is partitioned in 3, with 40 GB for RAN, 20 GB for LLM, and 10 GB for ResNet/CNN.}

\begin{figure}[t]
\centering
\setlength\fwidth{0.8\columnwidth} 
\setlength\fheight{0.35\columnwidth} 
\input{images/results/combined_medians_bars.tex}
\caption{\newrev{Median values for throughput and CRC error rate with RAN only (solid bars) or AI-RAN coexistence with LLM prompts and CNN/ResNet (dotted bars).}}
\label{fig:throughput}
\end{figure}

\newrev{Figure~\ref{fig:throughput} reports the median throughput and CRC error rate at the MAC layer, thus including retransmissions, with 95\% confidence intervals computed with the bootstrap method. The solid bars report the performance without AI; dotted bars are for experiments with \gls{ue} traffic, the CNN/ResNet training, and the LLM being prompted. The performance is consistent across all experiments, validating AI-RAN coexistence even in a single-node site.}

\newrev{Figure~\ref{fig:latency} illustrates the sequential deployment latency for different \glspl{llm} models and parameter sizes. We profile the workload deployment latency from when the AI-RAN site receives the request to when the service is available, assuming a sequential deployment for the LLM, followed by the RAN and ResNet. As can be seen, the \gls{ran} and ResNet are deployed in 1-4 s, with the longer time following the deployment of the largest Llama model that fits the MIG partition. \gls{llm} instances require between 5.8-34.4\:s (average). This shows that complex workloads can be managed on the same infrastructure in a matter of seconds, and provides an indication of the dynamicity with which such workloads can be (re)deployed by the AI-SMO.}

\begin{figure}[t]
\centering
\setlength\fwidth{0.95\columnwidth} 
\setlength\fheight{0.35\columnwidth} 
\input{images/results/deployment-latency.tex}
\caption{\newrev{Workload deployment latency on the AI-RAN site.}}
\label{fig:latency}
\end{figure}

\section{Conclusions}
\label{sec:conclusions}

We presented an end-to-end network architecture for AI-RAN coexistence. After reviewing the AI-RAN Alliance activities and O-RAN architecture, we discussed the challenges associated to operating AI and RAN on the same infrastructure. We then proposed an extension of the O-RAN architecture and O-RAN components that enables the deployment and management of AI workloads on the shared RAN infrastructure, including different modes for managing AI workloads with different requirements for deployment latency and location. \newrev{Near-term extensions of this work include further development of the architecture into an end-to-end prototype serving as a platform to validate resource allocation strategies. In the medium term, it will also facilitate studying how interfaces and design can be streamlined into O-RAN and AI-RAN Alliances activities and specifications.}

\vspace{-0.2cm}
\balance
\footnotesize  
\bibliographystyle{IEEEtran}
\bibliography{biblio.bib}

\vspace{-1.5cm}
\begin{IEEEbiographynophoto}
{Michele Polese} is a Research Assistant Professor at Northeastern. He received his Ph.D. from the University of Padova in 2020.

\noindent\textbf{Niloofar Mohamadi} is Principal Software Engineer at Northeastern. She received her Ph.D. from Ontario Tech University in 2022.

\noindent\textbf{Salvatore D'Oro} is a Research Associate Professor at Northeastern and the CTO at zTouch Networks. He received his Ph.D.\ from the University of Catania in 2015. 

\noindent\textbf{Leonardo Bonati} is an Associate Research Scientist at Northeastern University, where he received his Ph.D. in 2022. 

\noindent\textbf{Tommaso Melodia} is the William Lincoln Smith Chair Professor at Northeastern University and Director of the Institute for the Wireless Internet of Things.
\end{IEEEbiographynophoto}





\end{document}

%% file: acronyms.tex
\newacronym{3gpp}{3GPP}{3rd Generation Partnership Project}
\newacronym{4g}{4G}{4th generation}
\newacronym{5g}{5G}{5th generation}
\newacronym{6g}{6G}{6th generation}
\newacronym{5gc}{5GC}{5G Core}
\newacronym{adc}{ADC}{Analog to Digital Converter}
\newacronym{aerpaw}{AERPAW}{Aerial Experimentation and Research Platform for Advanced Wireless}
\newacronym{ai}{AI}{Artificial Intelligence}
\newacronym{aimd}{AIMD}{Additive Increase Multiplicative Decrease}
\newacronym{am}{AM}{Acknowledged Mode}
\newacronym{amc}{AMC}{Adaptive Modulation and Coding}
\newacronym{amf}{AMF}{Access and Mobility Management Function}
\newacronym{aops}{AOPS}{Adaptive Order Prediction Scheduling}
\newacronym{api}{API}{Application Programming Interface}
\newacronym{apn}{APN}{Access Point Name}
\newacronym{ap}{AP}{Application Protocol}
\newacronym{aqm}{AQM}{Active Queue Management}
\newacronym{ausf}{AUSF}{Authentication Server Function}
\newacronym{avc}{AVC}{Advanced Video Coding}
\newacronym{awgn}{AGWN}{Additive White Gaussian Noise}
\newacronym{balia}{BALIA}{Balanced Link Adaptation Algorithm}
\newacronym{bbu}{BBU}{Base Band Unit}
\newacronym{bdp}{BDP}{Bandwidth-Delay Product}
\newacronym{ber}{BER}{Bit Error Rate}
\newacronym{bf}{BF}{Beamforming}
\newacronym{bler}{BLER}{Block Error Rate}
\newacronym{brr}{BRR}{Bayesian Ridge Regressor}
\newacronym{bs}{BS}{Base Station}
\newacronym{bsr}{BSR}{Buffer Status Report}
\newacronym{bss}{BSS}{Business Support System}
\newacronym{ca}{CA}{Carrier Aggregation}
\newacronym{caas}{CaaS}{Connectivity-as-a-Service}
\newacronym{cb}{CB}{Code Block}
\newacronym{cc}{CC}{Congestion Control}
\newacronym{ccid}{CCID}{Congestion Control ID}
\newacronym{cco}{CC}{Carrier Component}
\newacronym{cdd}{CDD}{Cyclic Delay Diversity}
\newacronym{cdf}{CDF}{Cumulative Distribution Function}
\newacronym{cdn}{CDN}{Content Distribution Network}
\newacronym{cn}{CN}{Core Network}
\newacronym{codel}{CoDel}{Controlled Delay Management}
\newacronym{comac}{COMAC}{Converged Multi-Access and Core}
\newacronym{cord}{CORD}{Central Office Re-architected as a Datacenter}
\newacronym{cornet}{CORNET}{COgnitive Radio NETwork}
\newacronym{cosmos}{COSMOS}{Cloud Enhanced Open Software Defined Mobile Wireless Testbed for City-Scale Deployment}
\newacronym{cots}{COTS}{Commercial Off-the-Shelf}
\newacronym{cp}{CP}{Control Plane}
\newacronym{cyp}{CP}{Cyclic Prefix}
\newacronym{up}{UP}{User Plane}
\newacronym{cpu}{CPU}{Central Processing Unit}
\newacronym{cqi}{CQI}{Channel Quality Information}
\newacronym{cr}{CR}{Cognitive Radio}
\newacronym{cran}{CRAN}{Cloud \gls{ran}}
\newacronym{crs}{CRS}{Cell Reference Signal}
\newacronym{csi}{CSI}{Channel State Information}
\newacronym{csirs}{CSI-RS}{Channel State Information - Reference Signal}
\newacronym{cu}{CU}{Central Unit}
\newacronym{d2tcp}{D$^2$TCP}{Deadline-aware Data center TCP}
\newacronym{d3}{D$^3$}{Deadline-Driven Delivery}
\newacronym{dac}{DAC}{Digital to Analog Converter}
\newacronym{dag}{DAG}{Directed Acyclic Graph}
\newacronym{das}{DAS}{Distributed Antenna System}
\newacronym{dash}{DASH}{Dynamic Adaptive Streaming over HTTP}
\newacronym{dc}{DC}{Dual Connectivity}
\newacronym{dccp}{DCCP}{Datagram Congestion Control Protocol}
\newacronym{dce}{DCE}{Direct Code Execution}
\newacronym{dci}{DCI}{Downlink Control Information}
\newacronym{dctcp}{DCTCP}{Data Center TCP}
\newacronym{dl}{DL}{Deep Learning}
\newacronym{dmr}{DMR}{Deadline Miss Ratio}
\newacronym{dmrs}{DMRS}{DeModulation Reference Signal}
\newacronym{drlcc}{DRL-CC}{Deep Reinforcement Learning Congestion Control}
\newacronym{drs}{DRS}{Discovery Reference Signal}
\newacronym{du}{DU}{Distributed Unit}
\newacronym{e2e}{E2E}{end-to-end}
\newacronym{ecaas}{ECaaS}{Edge-Cloud-as-a-Service}
\newacronym{ecn}{ECN}{Explicit Congestion Notification}
\newacronym{edf}{EDF}{Earliest Deadline First}
\newacronym{embb}{eMBB}{Enhanced Mobile Broadband}
\newacronym{empower}{EMPOWER}{EMpowering transatlantic PlatfOrms for advanced WirEless Research}
\newacronym{enb}{eNB}{evolved Node Base}
\newacronym{endc}{EN-DC}{E-UTRAN-\gls{nr} \gls{dc}}
\newacronym{epc}{EPC}{Evolved Packet Core}
\newacronym{eps}{EPS}{Evolved Packet System}
\newacronym{es}{ES}{Edge Server}
\newacronym{etsi}{ETSI}{European Telecommunications Standards Institute}
\newacronym[firstplural=Estimated Times of Arrival (ETAs)]{eta}{ETA}{Estimated Time of Arrival}
\newacronym{eutran}{E-UTRAN}{Evolved Universal Terrestrial Access Network}
\newacronym{faas}{FaaS}{Function-as-a-Service}
\newacronym{fapi}{FAPI}{Functional Application Platform Interface}
\newacronym{fdd}{FDD}{Frequency Division Duplexing}
\newacronym{fdm}{FDM}{Frequency Division Multiplexing}
\newacronym{fdma}{FDMA}{Frequency Division Multiple Access}
\newacronym{fed4fire}{FED4FIRE+}{Federation 4 Future Internet Research and Experimentation Plus}
\newacronym{fir}{FIR}{Finite Impulse Response}
\newacronym{fit}{FIT}{Future \acrlong{iot}}
\newacronym{fpga}{FPGA}{Field Programmable Gate Array}
\newacronym{fr2}{FR2}{Frequency Range 2}
\newacronym{fs}{FS}{Fast Switching}
\newacronym{fscc}{FSCC}{Flow Sharing Congestion Control}
\newacronym{ftp}{FTP}{File Transfer Protocol}
\newacronym{fw}{FW}{Flow Window}
\newacronym{ge}{GE}{Gaussian Elimination}
\newacronym{gnb}{gNB}{Next Generation Node Base}
\newacronym{gop}{GOP}{Group of Pictures}
\newacronym{gpr}{GPR}{Gaussian Process Regressor}
\newacronym{gpu}{GPU}{Graphics Processing Unit}
\newacronym{gtp}{GTP}{GPRS Tunneling Protocol}
\newacronym{gtpc}{GTP-C}{GPRS Tunnelling Protocol Control Plane}
\newacronym{gtpu}{GTP-U}{GPRS Tunnelling Protocol User Plane}
\newacronym{gtpv2c}{GTPv2-C}{\gls{gtp} v2 - Control}
\newacronym{gw}{GW}{Gateway}
\newacronym{harq}{HARQ}{Hybrid Automatic Repeat reQuest}
\newacronym{hetnet}{HetNet}{Heterogeneous Network}
\newacronym{hh}{HH}{Hard Handover}
\newacronym{hol}{HOL}{Head-of-Line}
\newacronym{hqf}{HQF}{Highest-quality-first}
\newacronym{hss}{HSS}{Home Subscription Server}
\newacronym{http}{HTTP}{HyperText Transfer Protocol}
\newacronym{ia}{IA}{Initial Access}
\newacronym{iab}{IAB}{Integrated Access and Backhaul}
\newacronym{ic}{IC}{Incident Command}
\newacronym{ietf}{IETF}{Internet Engineering Task Force}
\newacronym{ims}{IMS}{Infrastructure Management Service}
\newacronym{imsi}{IMSI}{International Mobile Subscriber Identity}
\newacronym{imt}{IMT}{International Mobile Telecommunication}
\newacronym{iot}{IoT}{Internet of Things}
\newacronym{ip}{IP}{Internet Protocol}
\newacronym{itu}{ITU}{International Telecommunication Union}
\newacronym{kpi}{KPI}{Key Performance Indicator}
\newacronym{kpm}{KPM}{Key Performance Measurement}
\newacronym{kvm}{KVM}{Kernel-based Virtual Machine}
\newacronym{los}{LOS}{Line-of-Sight}
\newacronym{lsm}{LSM}{Link-to-System Mapping}
\newacronym{lstm}{LSTM}{Long Short Term Memory}
\newacronym{lte}{LTE}{Long Term Evolution}
\newacronym{lxc}{LXC}{Linux Container}
\newacronym{m2m}{M2M}{Machine to Machine}
\newacronym{mac}{MAC}{Medium Access Control}
\newacronym{manet}{MANET}{Mobile Ad Hoc Network}
\newacronym{mano}{MANO}{Management and Orchestration}
\newacronym{mc}{MC}{Multi-Connectivity}
\newacronym{mcc}{MCC}{Mobile Cloud Computing}
\newacronym{mchem}{MCHEM}{Massive Channel Emulator}
\newacronym{mcs}{MCS}{Modulation and Coding Scheme}
\newacronym{mec}{MEC}{Multi-access Edge Computing}
\newacronym{mec2}{MEC}{Mobile Edge Cloud}
\newacronym{mfc}{MFC}{Mobile Fog Computing}
\newacronym{mgen}{MGEN}{Multi-Generator}
\newacronym{mi}{MI}{Mutual Information}
\newacronym{mib}{MIB}{Master Information Block}
\newacronym{miesm}{MIESM}{Mutual Information Based Effective SINR}
\newacronym{mimo}{MIMO}{Multiple Input, Multiple Output}
\newacronym{ml}{ML}{Machine Learning}
\newacronym{mlr}{MLR}{Maximum-local-rate}
\newacronym[plural=\gls{mme}s,firstplural=Mobility Management Entities (MMEs)]{mme}{MME}{Mobility Management Entity}
\newacronym{mmtc}{mMTC}{Massive Machine-Type Communications}
\newacronym{mmwave}{mmWave}{millimeter wave}
\newacronym{mpdccp}{MP-DCCP}{Multipath Datagram Congestion Control Protocol}
\newacronym{mptcp}{MPTCP}{Multipath TCP}
\newacronym{mr}{MR}{Maximum Rate}
\newacronym{mrdc}{MR-DC}{Multi \gls{rat} \gls{dc}}
\newacronym{mse}{MSE}{Mean Square Error}
\newacronym{mss}{MSS}{Maximum Segment Size}
\newacronym{mt}{MT}{Mobile Termination}
\newacronym{mtd}{MTD}{Machine-Type Device}
\newacronym{mtu}{MTU}{Maximum Transmission Unit}
\newacronym{mumimo}{MU-MIMO}{Multi-user \gls{mimo}}
\newacronym{mvno}{MVNO}{Mobile Virtual Network Operator}
\newacronym{nalu}{NALU}{Network Abstraction Layer Unit}
\newacronym{nas}{NAS}{Non-Access Stratum}
\newacronym{nbiot}{NB-IoT}{Narrow Band IoT}
\newacronym{nfv}{NFV}{Network Function Virtualization}
\newacronym{nfvi}{NFVI}{Network Function Virtualization Infrastructure}
\newacronym{ni}{NI}{Network Interfaces}
\newacronym{nic}{NIC}{Network Interface Card}
\newacronym{nlos}{NLOS}{Non-Line-of-Sight}
\newacronym{now}{NOW}{Non Overlapping Window}
\newacronym{nsm}{NSM}{Network Service Mesh}
\newacronym[type=hidden]{nr}{NR}{New Radio}
\newacronym{nrf}{NRF}{Network Repository Function}
\newacronym{nsa}{NSA}{Non Stand Alone}
\newacronym{nse}{NSE}{Network Slicing Engine}
\newacronym{nssf}{NSSF}{Network Slice Selection Function}
\newacronym{o2i}{O2I}{Outdoor to Indoor}
\newacronym{oai}{OAI}{OpenAirInterface}
\newacronym{oaicn}{OAI-CN}{\gls{oai} \acrlong{cn}}
\newacronym{oairan}{OAI-RAN}{\acrlong{oai} \acrlong{ran}}
\newacronym{oam}{OAM}{Operations, Administration and Maintenance}
\newacronym{ofdm}{OFDM}{Orthogonal Frequency Division Multiplexing}
\newacronym{olia}{OLIA}{Opportunistic Linked Increase Algorithm}
\newacronym{omec}{OMEC}{Open Mobile Evolved Core}
\newacronym{onap}{ONAP}{Open Network Automation Platform}
\newacronym{onf}{ONF}{Open Networking Foundation}
\newacronym{onos}{ONOS}{Open Networking Operating System}
\newacronym{oom}{OOM}{\gls{onap} Operations Manager}
\newacronym{opnfv}{OPNFV}{Open Platform for \gls{nfv}}
\newacronym[type=hidden]{oran}{O-RAN}{O-RAN}
\newacronym{orbit}{ORBIT}{Open-Access Research Testbed for Next-Generation Wireless Networks}
\newacronym{os}{OS}{Operating System}
\newacronym{oss}{OSS}{Operations Support System}
\newacronym{pa}{PA}{Position-aware}
\newacronym{pase}{PASE}{Prioritization, Arbitration, and Self-adjusting Endpoints}
\newacronym{pawr}{PAWR}{Platforms for Advanced Wireless Research}
\newacronym{pbch}{PBCH}{Physical Broadcast Channel}
\newacronym{pcef}{PCEF}{Policy and Charging Enforcement Function}
\newacronym{pcfich}{PCFICH}{Physical Control Format Indicator Channel}
\newacronym{pcrf}{PCRF}{Policy and Charging Rules Function}
\newacronym{pdcch}{PDCCH}{Physical Downlink Control Channel}
\newacronym{pdcp}{PDCP}{Packet Data Convergence Protocol}
\newacronym{pdsch}{PDSCH}{Physical Downlink Shared Channel}
\newacronym{pdu}{PDU}{Packet Data Unit}
\newacronym{pf}{PF}{Proportional Fair}
\newacronym{pgw}{PGW}{Packet Gateway}
\newacronym{phich}{PHICH}{Physical Hybrid ARQ Indicator Channel}
\newacronym{phy}{PHY}{Physical}
\newacronym{pmch}{PMCH}{Physical Multicast Channel}
\newacronym{pmi}{PMI}{Precoding Matrix Indicators}
\newacronym{powder}{POWDER}{Platform for Open Wireless Data-driven Experimental Research}
\newacronym{ppo}{PPO}{Proximal Policy Optimization}
\newacronym{ppp}{PPP}{Poisson Point Process}
\newacronym{prach}{PRACH}{Physical Random Access Channel}
\newacronym{prb}{PRB}{Physical Resource Block}
\newacronym{psnr}{PSNR}{Peak Signal to Noise Ratio}
\newacronym{pss}{PSS}{Primary Synchronization Signal}
\newacronym{pucch}{PUCCH}{Physical Uplink Control Channel}
\newacronym{pusch}{PUSCH}{Physical Uplink Shared Channel}
\newacronym{qam}{QAM}{Quadrature Amplitude Modulation}
\newacronym{qci}{QCI}{\gls{qos} Class Identifier}
\newacronym{qoe}{QoE}{Quality of Experience}
\newacronym{qos}{QoS}{Quality of Service}
\newacronym{quic}{QUIC}{Quick UDP Internet Connections}
\newacronym{rach}{RACH}{Random Access Channel}
\newacronym{ran}{RAN}{Radio Access Network}
\newacronym[firstplural=Radio Access Technologies (RATs)]{rat}{RAT}{Radio Access Technology}
\newacronym{rcn}{RCN}{Research Coordination Network}
\newacronym{rc}{RC}{RAN Control}
\newacronym{rec}{REC}{Radio Edge Cloud}
\newacronym{red}{RED}{Random Early Detection}
\newacronym{renew}{RENEW}{Reconfigurable Eco-system for Next-generation End-to-end Wireless}
\newacronym{rf}{RF}{Radio Frequency}
\newacronym{rfc}{RFC}{Request for Comments}
\newacronym{rfr}{RFR}{Random Forest Regressor}
\newacronym{ric}{RIC}{RAN Intelligent Controller}
\newacronym{rlc}{RLC}{Radio Link Control}
\newacronym{rlf}{RLF}{Radio Link Failure}
\newacronym{rlnc}{RLNC}{Random Linear Network Coding}
\newacronym{rmr}{RMR}{RIC Message Router}
\newacronym{rmse}{RMSE}{Root Mean Squared Error}
\newacronym{rnis}{RNIS}{Radio Network Information Service}
\newacronym{rr}{RR}{Round Robin}
\newacronym{rrc}{RRC}{Radio Resource Control}
\newacronym{rrm}{RRM}{Radio Resource Management}
\newacronym{rru}{RRU}{Remote Radio Unit}
\newacronym{rs}{RS}{Remote Server}
\newacronym{rsrp}{RSRP}{Reference Signal Received Power}
\newacronym{rsrq}{RSRQ}{Reference Signal Received Quality}
\newacronym{rss}{RSS}{Received Signal Strength}
\newacronym{rssi}{RSSI}{Received Signal Strength Indicator}
\newacronym{rtt}{RTT}{Round Trip Time}
\newacronym{ru}{RU}{Radio Unit}
\newacronym{rw}{RW}{Receive Window}
\newacronym{rx}{RX}{Receiver}
\newacronym{s1ap}{S1AP}{S1 Application Protocol}
\newacronym{sa}{SA}{standalone}
\newacronym{sack}{SACK}{Selective Acknowledgment}
\newacronym{sap}{SAP}{Service Access Point}
\newacronym{sc2}{SC2}{Spectrum Collaboration Challenge}
\newacronym{scef}{SCEF}{Service Capability Exposure Function}
\newacronym{sch}{SCH}{Secondary Cell Handover}
\newacronym{scoot}{SCOOT}{Split Cycle Offset Optimization Technique}
\newacronym{sctp}{SCTP}{Stream Control Transmission Protocol}
\newacronym{sdap}{SDAP}{Service Data Adaptation Protocol}
\newacronym{sdk}{SDK}{Software Development Kit}
\newacronym{sdm}{SDM}{Space Division Multiplexing}
\newacronym{sdma}{SDMA}{Spatial Division Multiple Access}
\newacronym{sdn}{SDN}{Software-defined Networking}
\newacronym{sdr}{SDR}{Software-defined Radio}
\newacronym{seba}{SEBA}{SDN-Enabled Broadband Access}
\newacronym{sgsn}{SGSN}{Serving GPRS Support Node}
\newacronym{sgw}{SGW}{Service Gateway}
\newacronym{si}{SI}{Study Item}
\newacronym{sib}{SIB}{Secondary Information Block}
\newacronym{sinr}{SINR}{Signal to Interference plus Noise Ratio}
\newacronym{sip}{SIP}{Session Initiation Protocol}
\newacronym{siso}{SISO}{Single Input, Single Output}
\newacronym{sla}{SLA}{Service Level Agreement}
\newacronym{sm}{SM}{Service Model}
\newacronym{smf}{SMF}{Session Management Function}
\newacronym{smo}{SMO}{Service Management and Orchestration}
\newacronym{sms}{SMS}{Short Message Service}
\newacronym{smsgmsc}{SMS-GMSC}{\gls{sms}-Gateway}
\newacronym{snr}{SNR}{Signal-to-Noise-Ratio}
\newacronym{son}{SON}{Self-Organizing Network}
\newacronym{sptcp}{SPTCP}{Single Path TCP}
\newacronym{srb}{SRB}{Service Radio Bearer}
\newacronym{srn}{SRN}{Standard Radio Node}
\newacronym{srs}{SRS}{Sounding Reference Signal}
\newacronym{ss}{SS}{Synchronization Signal}
\newacronym{sss}{SSS}{Secondary Synchronization Signal}
\newacronym{st}{ST}{Spanning Tree}
\newacronym{svc}{SVC}{Scalable Video Coding}
\newacronym{tb}{TB}{Transport Block}
\newacronym{tcp}{TCP}{Transmission Control Protocol}
\newacronym{tdd}{TDD}{Time Division Duplexing}
\newacronym{tdm}{TDM}{Time Division Multiplexing}
\newacronym{tdma}{TDMA}{Time Division Multiple Access}
\newacronym{tfl}{TfL}{Transport for London}
\newacronym{tfrc}{TFRC}{TCP-Friendly Rate Control}
\newacronym{tft}{TFT}{Traffic Flow Template}
\newacronym{tgen}{TGEN}{Traffic Generator}
\newacronym{tip}{TIP}{Telecom Infra Project}
\newacronym{tm}{TM}{Transparent Mode}
\newacronym{to}{TO}{Telco Operator}
\newacronym{tr}{TR}{Technical Report}
\newacronym{trp}{TRP}{Transmitter Receiver Pair}
\newacronym{ts}{TS}{Technical Specification}
\newacronym{tti}{TTI}{Transmission Time Interval}
\newacronym{ttt}{TTT}{Time-to-Trigger}
\newacronym{tx}{TX}{Transmitter}
\newacronym{uas}{UAS}{Unmanned Aerial System}
\newacronym{uav}{UAV}{Unmanned Aerial Vehicle}
\newacronym{udm}{UDM}{Unified Data Management}
\newacronym{udp}{UDP}{User Datagram Protocol}
\newacronym{udr}{UDR}{Unified Data Repository}
\newacronym{ue}{UE}{User Equipment}
\newacronym{uhd}{UHD}{\gls{usrp} Hardware Driver}
\newacronym{ul}{UL}{Uplink}
\newacronym{um}{UM}{Unacknowledged Mode}
\newacronym{uml}{UML}{Unified Modeling Language}
\newacronym{upa}{UPA}{Uniform Planar Array}
\newacronym{upf}{UPF}{User Plane Function}
\newacronym{urllc}{URLLC}{Ultra Reliable and Low Latency Communications}
\newacronym{usa}{U.S.}{United States}
\newacronym{usim}{USIM}{Universal Subscriber Identity Module}
\newacronym{usrp}{USRP}{Universal Software Radio Peripheral}
\newacronym{utc}{UTC}{Urban Traffic Control}
\newacronym{vim}{VIM}{Virtualization Infrastructure Manager}
\newacronym{vm}{VM}{Virtual Machine}
\newacronym{vnf}{VNF}{Virtual Network Function}
\newacronym{volte}{VoLTE}{Voice over \gls{lte}}
\newacronym{voltha}{VOLTHA}{Virtual OLT HArdware Abstraction}
\newacronym{vr}{VR}{Virtual Reality}
\newacronym{vran}{vRAN}{Virtualized \gls{ran}}
\newacronym{vss}{VSS}{Video Streaming Server}
\newacronym{wbf}{WBF}{Wired Bias Function}
\newacronym{wf}{WF}{Waterfilling}
\newacronym{wg}{WG}{Working Group}
\newacronym{wlan}{WLAN}{Wireless Local Area Network}
\newacronym{osm}{OSM}{Open Source \gls{nfv} Management and Orchestration}
\newacronym{pnf}{PNF}{Physical Network Function}
\newacronym{drl}{DRL}{Deep Reinforcement Learning}
\newacronym{mtc}{MTC}{Machine-type Communications}
\newacronym{osc}{OSC}{O-RAN Software Community}
\newacronym{mns}{MnS}{Management Services}
\newacronym{ves}{VES}{\gls{vnf} Event Stream}
\newacronym{ei}{EI}{Enrichment Information}
\newacronym{fh}{FH}{Fronthaul}
\newacronym{fft}{FFT}{Fast Fourier Transform}
\newacronym{laa}{LAA}{Licensed-Assisted Access}
\newacronym{plfs}{PLFS}{Physical Layer Frequency Signals}
\newacronym{ptp}{PTP}{Precision Time Protocol}
\newacronym{cnn}{CNN}{Convolutional Neural Network}
\newacronym{aoa}{AoA}{Angle of Arrival}
\newacronym{xr}{XR}{Extended Reality}
\newacronym{icc}{ICC}{Intelligence Coordination Controller}
\newacronym{smos}{SMOS}{SMO Services}
\newacronym{focom}{FOCOM}{Federated O-Cloud Orchestration and Management}
\newacronym{nfo}{NFO}{Network Function Orchestration}
\newacronym{dms}{DMS}{Deployment Management Service}
\newacronym{llm}{LLM}{Large Language Model}
\newacronym{isac}{ISAC}{Integrated Sensing and Communications}
\newacronym{mig}{MIG}{Multi-Instance GPU}
\newacronym{dsp}{DSP}{Digital Signal Processing}

%% file: myTikz.tex
\tikzstyle{startstop} = [rectangle, rounded corners, minimum width=2cm, minimum height=0.5cm,text centered, draw=black]
\tikzstyle{io} = [trapezium, trapezium left angle=70, trapezium right angle=110, minimum width=3cm, minimum height=1cm, text centered, draw=black]
\tikzstyle{process} = [rectangle, minimum width=2cm, minimum height=0.5cm, text centered, draw=black, alignb=center]
\tikzstyle{decision} = [ellipse, minimum width=2cm, minimum height=1cm, text centered, draw=black]
\tikzstyle{arrow} = [thick,<->,>=stealth]
\tikzstyle{line} = [thick,>=stealth]
\tikzstyle{darrow} = [thick,<->,>=stealth,dashed]
\tikzstyle{sarrow} = [thick,->,>=stealth]
\tikzstyle{larrow} = [line width=0.1mm,dashdotted,->,>=stealth]
\tikzstyle{llarrow} = [line width=0.1mm,->,>=stealth]

\makeatletter
\def\grd@save@target#1{%
  \def\grd@target{#1}}
\def\grd@save@start#1{%
  \def\grd@start{#1}}
\tikzset{
  grid with coordinates/.style={
    to path={%
      \pgfextra{%
        \edef\grd@@target{(\tikztotarget)}%
        \tikz@scan@one@point\grd@save@target\grd@@target\relax
        \edef\grd@@start{(\tikztostart)}%
        \tikz@scan@one@point\grd@save@start\grd@@start\relax
        \draw[minor help lines] (\tikztostart) grid (\tikztotarget);
        \draw[major help lines] (\tikztostart) grid (\tikztotarget);
        \grd@start
        \pgfmathsetmacro{\grd@xa}{\the\pgf@x/1cm}
        \pgfmathsetmacro{\grd@ya}{\the\pgf@y/1cm}
        \grd@target
        \pgfmathsetmacro{\grd@xb}{\the\pgf@x/1cm}
        \pgfmathsetmacro{\grd@yb}{\the\pgf@y/1cm}
        \pgfmathsetmacro{\grd@xc}{\grd@xa + \pgfkeysvalueof{/tikz/grid with coordinates/major step x}}
        \pgfmathsetmacro{\grd@yc}{\grd@ya + \pgfkeysvalueof{/tikz/grid with coordinates/major step y}}
        \foreach \x in {\grd@xa,\grd@xc,...,\grd@xb}
        \node[anchor=north] at (\x,\grd@ya) {\pgfmathprintnumber{\x}};
        \foreach \y in {\grd@ya,\grd@yc,...,\grd@yb}
        \node[anchor=east] at (\grd@xa,\y) {\pgfmathprintnumber{\y}};
      }
    }
  },
  minor help lines/.style={
    help lines,
    gray,
    line cap =round,
    xstep=\pgfkeysvalueof{/tikz/grid with coordinates/minor step x},
    ystep=\pgfkeysvalueof{/tikz/grid with coordinates/minor step y}
  },
  major help lines/.style={
    help lines,
    line cap =round,
    line width=\pgfkeysvalueof{/tikz/grid with coordinates/major line width},
    xstep=\pgfkeysvalueof{/tikz/grid with coordinates/major step x},
    ystep=\pgfkeysvalueof{/tikz/grid with coordinates/major step y}
  },
  grid with coordinates/.cd,
  minor step x/.initial=.5,
  minor step y/.initial=.2,
  major step x/.initial=1,
  major step y/.initial=1,
  major line width/.initial=1pt,
}
\makeatother

%% file: images/results/combined_medians_bars.tex
\definecolor{mycolor1}{rgb}{0.00000,0.44700,0.74100}%
\definecolor{mycolor2}{rgb}{0.85000,0.32500,0.09800}%
\begin{tikzpicture}
\pgfplotsset{every y tick label/.append style={font=\tiny}}
\pgfplotsset{every x tick label/.append style={font=\scriptsize}}

\begin{axis}[%
ybar,
width=0.951\fwidth,
height=\fheight,
at={(0\fwidth,0\fheight)},
scale only axis,
bar shift auto,
tick align=inside,
enlarge x limits=0.2,
xtick=data,
xticklabels={{CNN}, {ResNet}, {CNN}, {ResNet}},
xlabel style={font=\scriptsize\color{white!15!black}},
ymin=0,
ymax=65,
ylabel style={font=\scriptsize\color{mycolor1}},
ylabel={MAC Throughput [Mbps]},
axis background/.style={fill=white},
title style={font=\bfseries},
xmajorgrids,
ymajorgrids,
legend style={at={(0.5,1.141)}, anchor=south, legend cell align=left, align=left, draw=white!15!black, font=\tiny, legend columns=2},
y axis line style={color=mycolor1},     
ytick style={color=mycolor1},           
yticklabel style={color=mycolor1},      
ylabel shift=-5pt,
bar width=7pt
]

\node[anchor=south, font=\scriptsize] at (axis cs:1.5,57) {\textit{ ------ Llama-3.2-3B ------ }};
\node[anchor=south, font=\scriptsize] at (axis cs:3.5,57) {\textit{ ------ Llama-3-8B ------ }};

\addplot +[
	xshift=-.35cm,
    fill=mycolor1, 
    draw=black, 
    area legend,
    error bars/.cd,
    y dir=both,
    y explicit
] table[row sep=crcr, y error plus expr=\thisrow{eplus}-\thisrow{y}, y error minus expr=\thisrow{y}-\thisrow{eminus}] {%
x   y       eplus	eminus\\
1   55.96     57.3805	54.985\\
2   55.32     57.165	54.205\\
3   54.075    54.69		53.78\\
4   54.975    56.02		54.39\\
};
\addlegendentry{DL Th / RAN-only}

\addplot +[
	xshift=-.35cm,
    fill=mycolor1, 
    draw=black, 
    area legend,
    postaction={pattern=crosshatch dots},
    error bars/.cd,
    y dir=both,
    y explicit,
] table[row sep=crcr, y error plus expr=\thisrow{eplus}-\thisrow{y}, y error minus expr=\thisrow{y}-\thisrow{eminus}] {%
x   y       eminus	eplus\\
1   56.72     56		58.58\\
2   55.38     55.16		55.74\\
3   55.41    54.935		55.94\\
4   57.33    56.66		57.945\\
};
\addlegendentry{DL Th / AI-RAN Coexistence}

\end{axis}

\begin{axis}[%
ybar,
width=0.951\fwidth,
height=\fheight,
at={(0\fwidth,0\fheight)},
scale only axis,
bar shift auto,
ytick style={draw=none},
enlarge x limits=0.2,
ymin=0,
ymax=18,
ylabel style={font=\scriptsize\color{mycolor2}},
ylabel={CRC Error Rate [error/s]},
legend style={at={(0.5,1.01)}, anchor=south, legend cell align=left, align=left, draw=white!15!black, font=\tiny, legend columns=2},
axis y line*=right,  
axis x line=none,    
y axis line style={color=mycolor2},     
    ytick style={color=mycolor2},           
    yticklabel style={color=mycolor2},      
ylabel shift=-5pt,
bar width=7pt
]
\addplot +[
	xshift=+.35cm,
    fill=mycolor2, 
    draw=black, 
    area legend,
    error bars/.cd,
    y dir=both,
    y explicit,
] table[row sep=crcr, y error plus expr=\thisrow{eplus}-\thisrow{y}, y error minus expr=\thisrow{y}-\thisrow{eminus}] {%
x   y       eminus	eplus\\
1   8    4.5	13\\
2   9    3.5	11.5\\
3   5.5    3.5	13\\
4   9.5    5	13.5\\
};
\addlegendentry{CRC Error Rate / RAN-only}

\addplot +[
	xshift=+.35cm,
    fill=mycolor2, 
    draw=black, 
    area legend,
    postaction={pattern=crosshatch dots},
    error bars/.cd,
    y dir=both,
    y explicit,
] table[row sep=crcr, y error plus expr=\thisrow{eplus}-\thisrow{y}, y error minus expr=\thisrow{y}-\thisrow{eminus}] {%
x   y       eminus	eplus\\
1   7    5	9\\
2   8    6	9\\
3   8    6	10\\
4   7    5	9\\
};
\addlegendentry{CRC Error Rate / AI-RAN Coexistence}
\end{axis}

\end{tikzpicture}

%% file: images/results/deployment-latency.tex




\usepgfplotslibrary{groupplots}
\usepgfplotslibrary{colormaps}
\usetikzlibrary{arrows.meta,patterns,patterns.meta}
\tikzset{draw-color/.style={
		color of colormap={#1},
		draw=.!80,
	},
	fill-color/.style={
		color of colormap={#1},
		draw=.!80!black,
		fill=.!80!white,
		fill opacity=0.6
	},
	mydashed/.style={dash pattern=on 6pt off 4pt}
}

\begin{tikzpicture}

\definecolor{coral}{RGB}{255,127,80}
\definecolor{cornflowerblue}{RGB}{100,149,237}
\definecolor{darkgray176}{RGB}{176,176,176}
\definecolor{darkorchid}{RGB}{153,50,204}
\definecolor{skyblue}{RGB}{135,206,235}
\definecolor{mycolor6}{rgb}{0.30100,0.74500,0.93300}
\definecolor{mycolor5}{rgb}{0.46600,0.67400,0.18800}%
\definecolor{mycolor3}{rgb}{0.92900,0.19400,0.12500}%

\begin{axis}[
width=\fwidth,
height=\fheight,
xlabel={Time [s]},
xmin=0, xmax=43,
ymin=-0.24, ymax=4.04,
ytick={0.4,1.4,2.4,3.4},
xmajorgrids,
ymajorgrids,
yticklabels={
  {Qwen2.5-1.5B},
  {Llama-3.2-1B},
  {Llama-3.2-3B},
  {Llama-3-8B},
},
legend style={font=\scriptsize, at={(0.99,0.01)}, anchor=south east, legend columns=2},
tick label style={font=\scriptsize, align=right},
label style={font=\scriptsize},
xlabel shift=-4pt,
]



\path [fill=mycolor6]
(axis cs:0,0)
--(axis cs:0,0.8)
--(axis cs:14.338,0.8)
--(axis cs:14.338,0)
--(axis cs:0,0)
--cycle;

\path [fill=mycolor5,postaction={pattern=north east lines, pattern color=black}]
(axis cs:14.338,0)
--(axis cs:14.338,0.8)
--(axis cs:17.338,0.8)
--(axis cs:17.338,0)
--(axis cs:14.338,0)
--cycle;

\path [fill=mycolor3,postaction={pattern=vertical lines, pattern color=black}]
(axis cs:17.338,0)
--(axis cs:17.338,0.8)
--(axis cs:19.538,0.8)
--(axis cs:19.538,0)
--(axis cs:17.338,0)
--cycle;



\path [fill=mycolor6]
(axis cs:0,1)
--(axis cs:0,1.8)
--(axis cs:5.818,1.8)
--(axis cs:5.818,1)
--(axis cs:0,1)
--cycle;

\path [fill=mycolor5,postaction={pattern=north east lines, pattern color=black}]
(axis cs:5.818,1)
--(axis cs:5.818,1.8)
--(axis cs:9.018,1.8)
--(axis cs:9.018,1)
--(axis cs:5.818,1)
--cycle;

\path [fill=mycolor3,postaction={pattern=vertical lines, pattern color=black}]
(axis cs:9.018,1)
--(axis cs:9.018,1.8)
--(axis cs:10.818,1.8)
--(axis cs:10.818,1)
--(axis cs:9.018,1)
--cycle;


\path [fill=mycolor6]
(axis cs:0,2)
--(axis cs:0,2.8)
--(axis cs:9.702,2.8)
--(axis cs:9.702,2)
--(axis cs:0,2)
--cycle;

\path [fill=mycolor5,postaction={pattern=north east lines, pattern color=black}]
(axis cs:9.702,2)
--(axis cs:9.702,2.8)
--(axis cs:12.902,2.8)
--(axis cs:12.902,2)
--(axis cs:9.702,2)
--cycle;

\path [fill=mycolor3,postaction={pattern=vertical lines, pattern color=black}]
(axis cs:12.902,2)
--(axis cs:12.902,2.8)
--(axis cs:14.902,2.8)
--(axis cs:14.902,2)
--(axis cs:12.902,2)
--cycle;


\path [fill=mycolor6]
(axis cs:0,3)
--(axis cs:0,3.8)
--(axis cs:34.422,3.8)
--(axis cs:34.422,3)
--(axis cs:0,3)
--cycle;

\path [fill=mycolor5,postaction={pattern=north east lines, pattern color=black}]
(axis cs:34.422,3)
--(axis cs:34.422,3.8)
--(axis cs:39.022,3.8)
--(axis cs:39.022,3)
--(axis cs:34.422,3)
--cycle;

\path [fill=mycolor3,postaction={pattern=vertical lines, pattern color=black}]
(axis cs:39.022,3)
--(axis cs:39.022,3.8)
--(axis cs:42.022,3.8)
--(axis cs:42.022,3)
--(axis cs:39.022,3)
--cycle;

\addlegendimage{fill=mycolor6,area legend}
\addlegendentry{LLM}

\addlegendimage{fill=mycolor5,postaction={pattern=north east lines, pattern color=black},area legend}
\addlegendentry{RAN}

\addlegendimage{fill=mycolor3,postaction={pattern=vertical lines, pattern color=black},area legend}
\addlegendentry{CNN}

\end{axis}

\end{tikzpicture}